\documentclass[letterpaper]{article} 
\usepackage{aaai2026}  
\usepackage{times}  
\usepackage{helvet}  
\usepackage{courier}  
\usepackage[hyphens]{url}  
\usepackage{graphicx} 
\usepackage{natbib}  
\usepackage{caption} 
\usepackage{algorithm}
\usepackage{algorithmic}
\usepackage{multirow}
\usepackage{amsmath}
\usepackage{booktabs}
\usepackage{tabularx}
\usepackage{array}
\usepackage{newfloat}
\usepackage{listings}
\DeclareCaptionStyle{ruled}{labelfont=normalfont,labelsep=colon,strut=off} 
\floatstyle{ruled}
\newfloat{listing}{tb}{lst}{}
\floatname{listing}{Listing}
\nocopyright

\title{Seeking Help, Facing Harm: Auditing TikTok's Mental Health Recommendations}
\author {
    Pooriya Jamie,
    Amir Ghasemian,
    Homa Hosseinmardi
}
\affiliations {
    OASIS Lab, University of California, Los Angeles (UCLA), USA \\ 
    \{pjamie,amirgh,homahm\}@ucla.edu
}

\begin{document}

\maketitle

\begin{abstract}
Recommender systems on social media increasingly mediate how users encounter mental health content, yet it remains unclear whether they distinguish help-seeking from distress expression. We conduct a controlled 7-day audit of TikTok's ``For You'' page using 30 fresh accounts and LLM-guided agents that vary initial search framing (distress- vs.\ help-initiated) and interaction strategy (engaged, avoidant, passive). Across 8,727 recommended videos, interaction behavior dominates exposure outcomes: engagement rapidly saturates feeds with mental health content ($\simeq$45\% of daily recommendations), while avoidance and passive viewing reduce but do not eliminate exposure ($\simeq$11--20\%). Search framing mainly shifts composition rather than volume---help-initiated searches yield more potentially supportive material, yet potentially harmful content persists at low but non-zero levels, including content in the \textit{Suicide/Self-Harm} category. These findings suggest limited sensitivity to user intent signals in TikTok's recommendations and motivate context-aware safeguards for sensitive topics.
\end{abstract}

\section{Introduction}

Social media platforms have increasingly become spaces where users seek information and emotional support related to mental health~\citep{Naslund2016future, Akhther2022SeekingSharing, Saha2020Causal}. For individuals experiencing loneliness, depression, or emotional distress, algorithmically curated feeds can play a consequential role in shaping exposure to both supportive and harmful content~\citep{Milton2023ISee, Gillian2025Self, Ribeiro2023Amplification, Nguyen2025Supporters}. While recommender systems are optimized to personalize content based on user behavior, this optimization may inadvertently create feedback loops in which transient expressions of distress are interpreted as sustained interest, amplifying potentially harmful material~\citep{mansoury2020feedback,narayanan2023understanding,krauth2025breaking}.

These feedback loops are driven largely by implicit engagement signals. On TikTok, retention is weighted comparably to explicit signals such as likes or follows \citep{Boeker2022Empirical}, and more recent work indicates that watch time is now the primary driver of feed composition \citep{Mosnar2025Revisiting}. 
This personalization occurs rapidly: distinct filter bubbles can form within the first 200 videos of a session \citep{Baumann2025Dynamics}, narrowing the user's content reality.

These risks are especially acute for vulnerable populations, including minors and individuals experiencing depression---who may be susceptible to transitions toward suicidal ideation \citep{de2016discovering, Franklin2017RiskFactors, Nesi2021SocialMediaSITB}---yet may receive insufficient algorithmic protection. While methods exist to mitigate feedback loop bias in recommender systems \citep{krauth2025breaking}, it is not clear to what extent platforms have adopted such safeguards for vulnerable users. Audits of TikTok's age-gating features found that accounts registered as ``Youth'' (under 18) are exposed to harmful content at rates nearly identical to ``Adult'' accounts~\citep{Xue2025Towards}. In some contexts, minors are paradoxically exposed to higher frequencies of harmful content than adults~\citep{Eltaher2025Protecting}. Nor can users rely on transparency tools to navigate these risks, as platform explanations often fail to accurately reflect the behavioral reasons behind recommendations \citep{Mousavi2024Auditing}. Among these vulnerable populations, individuals seeking mental health support online face particular risks: as we show, the line between consuming distress-related content and seeking recovery is algorithmically ambiguous.

In this work, we focus on mental health content exposure on TikTok. While prior research established that algorithms generally exploit user interests to populate feeds while retaining global popularity signals \citep{Vombatkere2024TikTok}, a critical gap remains regarding the system's sensitivity to user intent. Specifically, it is unclear whether the algorithm distinguishes distress expression from help-seeking or treats both as a single undifferentiated mental health interest cluster. This distinction matters: conflating recovery-oriented intent with distress consumption may expose vulnerable users to content that exacerbates harm rather than mitigates it.

To address this gap, we conduct a controlled audit of TikTok's ``For You'' page (FYP) to examine how initial search framing and subsequent interaction behavior jointly shape mental health content exposure over time. We deploy LLM-driven simulated agents that semantically interpret content and make adaptive watch or skip decisions, enabling realistic yet controlled auditing of recommendation dynamics.

We focus specifically on the transient phase of algorithmic personalization, as early exposure patterns are especially consequential for vulnerable users. Prior research shows that warning signs of suicidal crisis emerge in social media weeks before an attempt \citep{coppersmith2016exploratory}, and that shifts from mental health content to suicidal ideation are detectable \citep{de2016discovering}. Understanding how recommendation systems behave during initial sessions is therefore critical for identifying early-stage harms. 

We vary two dimensions: initial search framing (distress expression vs. help-seeking) and behavioral interaction (engagement, avoidance, or passive observation). Crossing these dimensions yields six experimental conditions, allowing us to isolate the effects of stated intent (search framing) from revealed behavior (interaction patterns). Using this design, we address two research questions:

\noindent\textbf{RQ1}: 
How sensitive is TikTok's recommendation system to user intent signals conveyed through search and engagement? Does it distinguish between users who express distress and those who explicitly seek help, or are both treated as equivalent mental health interests?

\noindent\textbf{RQ2}: When intent-based safeguards are absent or ineffective, what harmful content are vulnerable users exposed to? Does exposure vary by user behavior (engagement, avoidance, passive observation)?

Our findings reveal that behavioral interaction is the dominant driver of algorithmic outcomes: users who engage with mental health content experience rapid feed saturation regardless of whether their initial search framing expressed distress or sought help, while those who actively avoid such content substantially suppress their exposure to mental health content---though not to zero---within the one-week study period. Most critically, we identify a help-seeking paradox: users who search for help and engage with mental health content receive more supportive material in the short term, but remain exposed to substantial potentially harmful content, including suicide/self-harm material. 
These results indicate that expressed intent through initial search framing does not meaningfully shape safety outcomes. 
The algorithm treats help-seeking and distress expression as equivalent signals, bundling supportive and potentially harmful content in its recommendations together and placing the burden of safety on users themselves.

\begin{figure*}[t]
    \centering
     
    \includegraphics[width=1\linewidth]{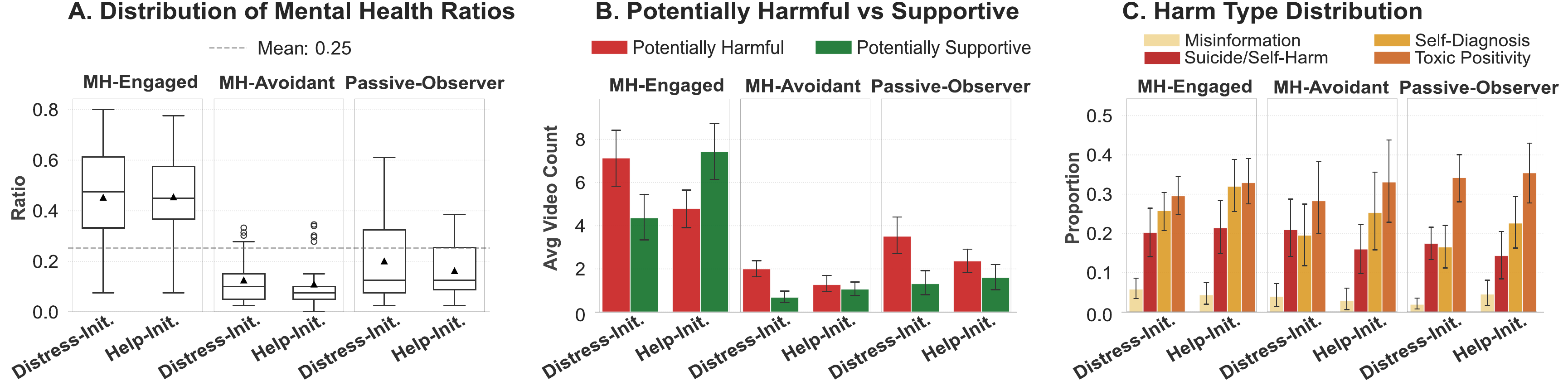}
    \caption{\textbf{Cross-sectional exposure patterns across personas.}
    (A) MH saturation (proportion of MH-relevant videos among 40 daily recommendations; dashed line is the global mean). (B) Average daily counts of potentially harmful vs.\ potentially supportive content. (C) Harm subtype proportions within potentially harmful content.}
    \label{fig:main_results}
\end{figure*}

\section{Audit Design and Experimental Setup}
We employed a 2$\times$3 factorial design, crossing initial search framing (distress-initiated [Distress-Init.] vs. help-initiated [Help-Init.]) with behavioral interaction patterns toward mental health (MH) content (MH-Engaged, MH-Avoidant, Passive-Observer). 
We created 30 accounts as ``fresh'' users with no prior history. Each account used an isolated device profile (distinct device identifiers) and a U.S. IP address to reduce cross-contamination. Agents interacted with TikTok (5 accounts per condition) over a 7-day period from January 1st to 7th, 2026, collecting a total of 8,727 videos from FYP. 

\textbf{Initial search framing.} On the first day, each agent executed a set of seed queries to establish baseline interest.
Distress-Init. agents searched with the following queries to simulate users who express emotional distress without explicitly requesting assistance: ``\textit{I feel depressed and lonely}'' and ``\textit{why do I feel so lonely and empty?}'' Help-Init. agents searched with the following queries to simulate users who seek support or guidance: ``\textit{depression help}'' and  ``\textit{mental health advice}.'' This distinction is consistent with prior observations that online mental health discourse includes both self-disclosure and help-seeking behaviors~\citep{Choudhury2014Mental}.
Searches were conducted only on day one. Agents watched 7 videos after each search (14 total), with randomized viewing durations between 10 and 15 seconds.

\textbf{Behavioral interaction.} Following the initial search phase, agents interacted with the FYP using a Multi-Modal Large Language Model (MLLM) to simulate distinct patterns of implicit engagement. Interactions differed only in how agents allocated watch time to videos identified as relevant to mental health, without using explicit feedback mechanisms such as likes or follows, as watch time is the primary driver of TikTok's recommendation algorithm \citep{Mosnar2025Revisiting}. Each agent scrolled through 40 videos per day. On day 1, scrolling began immediately after the search; from days 2--7, agents started directly from the FYP. 

Video content was classified in real-time using GPT-4o-mini with a structured prompt designed to identify mental health-relevant content, based on captions, hashtags, on-screen text, and visual elements. We defined mental health-relevant content as videos explicitly addressing mental health topics, emotional distress, coping strategies, or psychological well-being. Each video begins with a baseline viewing duration of $t_{\text{base}}$ ($\simeq$7 seconds), which supports screenshot capture and controller inference, and also serves as a proxy for the time users need to decide whether to continue watching or scroll. Based on the content type, agents dynamically either extended their watch time or skipped to the next video in accordance with their interaction strategy. Three behavioral interaction patterns were implemented: \textbf{MH-Engaged (strong reinforcement)} (watch MH-relevant videos for 25 additional seconds beyond $t_{\text{base}}$; skip non-relevant), \textbf{MH-Avoidant (active avoidance)} (skip MH-relevant videos after $t_{\text{base}}$; watch non-relevant for 25 additional seconds), and \textbf{Passive-Observer (neutral)} (watch all videos for 6--8 seconds regardless of content).
The 25-second extension signals interest. Similarly, after a few pilot tests, the choice of 40 videos was selected to balance ecological validity with data collection feasibility.
Operational reliability is reported in Appendix, section~C (Table~\ref{tab:ops}).

\textbf{Content labeling and taxonomy.}
The real-time classification described above served as a first-stage assessment to guide agent behavior by identifying MH-relevant videos. These videos then underwent post-hoc second-stage labeling, using textual metadata captured during the first stage, to categorize content by stance and potential impact.

MH-relevant videos were annotated into two categories \citep{Gillian2025Self}: ($i$) \textit{Potentially Supportive} content, containing professional advice, evidence-based coping strategies, and crisis support resources; and ($ii$) \textit{Potentially Harmful} content, which could normalize harmful behaviors, propagate misinformation, or exacerbate distress. Potentially harmful videos were further labeled by subtype, including Suicide/Self-Harm, Toxic Positivity, Self-Diagnosis, and Misinformation. We also flagged critical safety indicators, including explicit suicidal ideation, depiction of self-harm behaviors, and crisis helpline information.

We validated LLM-based classification performance using a random sample of 100 videos (approximately 1\% of the corpus), independently annotated by a human annotator using the same taxonomy. For MH-relevance detection, the LLM achieved F1 score of 77.42\%; for potentially harmful content detection, 64.52\%; for potentially supportive content, 64.00\% (see Tables~\ref{tab:prompt-schema}--\ref{tab:ablation_mh} for more details).  Because we compare conditions using the same labeling pipeline, we focus on between-condition differences, which should be relatively insensitive to moderate labeling noise when error rates are broadly similar across conditions.

\textbf{Experimental personas.} Crossing initial search framing with behavioral interaction yields six experimental personas: Distress-Init./MH-Engaged, Distress-Init./MH-Avoidant, Distress-Init./Passive-Observer, Help-Init./MH-Engaged, Help-Init./MH-Avoidant, Help-Init./Passive-Observer. This design allows us to disentangle the effects of stated intent (search framing) from revealed behavior (interaction patterns) on algorithmic content exposure.

\begin{figure*}[h]
    \centering
    \includegraphics[width=1\linewidth]{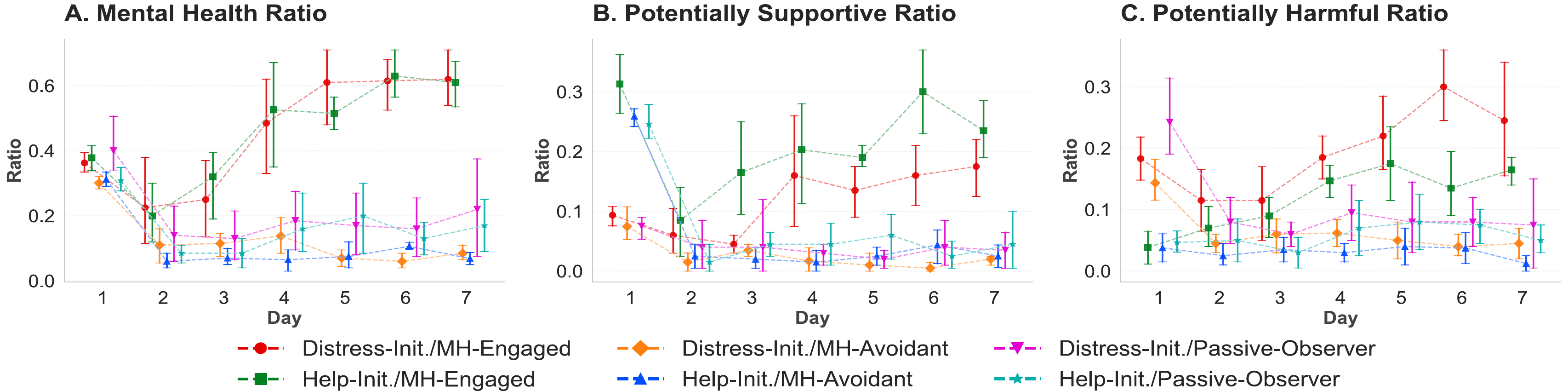}
\caption{\textbf{Feed evolution (Days 1--7).} (A) MH ratio over time by persona. MH-Engaged saturates while MH-Avoidant and Passive-Observer drop after Day 1 but remain non-zero. (B) Potentially supportive and (C) potentially harmful ratios, with harmful exposure persisting across conditions. Error bars show 95\% CIs.}
    \label{fig:temporal}
\end{figure*}

\section{Results}

Behavioral interaction plays a larger role than initial search framing in shaping MH content exposure. Across both cross-sectional summaries (Figure \ref{fig:main_results}) and 7-day trajectories (Figure \ref{fig:temporal}), MH-Engaged personas rapidly saturated their feeds, while MH-Avoidant and Passive-Observer personas experienced sharp declines in MH exposure after the initial search day. In what follows, we report MH saturation (the proportion of MH-relevant videos among the 40 videos shown per day) and, within MH content, the balance between potentially supportive and potentially harmful recommendations.

\noindent\textbf{Behavioral interaction signals dominate exposure dynamics.}
Across interaction strategies, MH-Engaged personas consistently experienced the highest levels of MH saturation, reaching approximately 45\% of their feed under both Distress-Init. and Help-Init. searches (Figure \ref{fig:main_results}A). In contrast, MH-Avoidant personas reduced MH exposure to roughly 11--13\%, while Passive-Observer personas remained at intermediate but closer-to-avoidant levels ($\simeq$16--20\%). These differences are substantially larger than those attributable to initial search framing, indicating that post-search behavior, rather than stated intent, is the primary separator of feed outcomes  
(see Table~\ref{tab:descriptives} for more details). 
Temporal trajectories reinforce this pattern. On Day 1 following the search, all personas start at comparable MH exposure levels. Afterward, MH-Engaged personas continue to accumulate MH content over several days before stabilizing at high saturation levels, whereas MH-Avoidant and Passive-Observer personas show a sharp drop on Day 2, after which MH exposure remains low and stable for the remainder of the experiment without fully disappearing (Figure \ref{fig:temporal}A). Notably, the persistence of MH exposure after Day 2 indicates that a single MH-related search leaves a lasting footprint even in the absence of continued engagement.

\noindent\textbf{Help-seeking shifts composition but does not protect users from risk.} While initial search framing has little effect on total MH volume, it more clearly influences the composition of MH content. Under MH-Engaged behavior, distress-initiated personas are exposed to a higher volume of potentially harmful than supportive content, whereas help-initiated personas show the opposite pattern, receiving more supportive material on average (Figure \ref{fig:main_results}B). This compositional shift is also visible longitudinally: supportive content grows more strongly over time for help-initiated MH-Engaged personas, while potentially harmful content rises most sharply for distress-initiated ones (Figure \ref{fig:temporal}B--C).

However, help-seeking does not act as a safety boundary. Across all behavioral interaction personas, potentially harmful content persists at non-zero levels, including for those who explicitly searched for help. Even among MH-Engaged help-initiated personas (with the highest volumes of supportive content) potentially harmful material continues to appear alongside it, indicating that the algorithm broadens exposure within the mental health topic but acts context-agnostically and at the cost of including potentially harmful material.

\noindent\textbf{Avoidance reduces volume but not risk.}
Actively avoiding MH content substantially lowers overall exposure, but does not eliminate potentially harmful material. For both MH-Avoidant and Passive-Observer personas, potentially harmful videos continue to appear at comparable proportions within the reduced MH feed during the 1-week experiment (Figures \ref{fig:main_results}B and \ref{fig:temporal}C). This suggests that avoidance suppresses how much MH content is shown, but does not reliably improve what that content contains.

\noindent\textbf{Harmful content is not confined to a single subtype.}
Across all personas, Toxic Positivity and Self-Diagnosis account for the largest shares of potentially harmful content, while highly sensitive content concerning suicide and self-harm also appears at slightly lower rates (Figure~\ref{fig:main_results}C). Differences in harmful subtype proportions across interaction strategies and search framings are comparatively modest, and no single risk subtype is confined to a specific persona. In the mental health context, this pattern indicates that risk exposure is not limited to extreme categories, as repeated exposure to subtler forms may also shape how users interpret symptoms and navigate support-seeking.
See Appendix, section ~D for supplementary results (Figure \ref{fig:suppfig}).

\section{Limitations and Ongoing Work}
This is ongoing work with clear limits. First, classifier performance is moderate, and errors in earlier stages may propagate through the pipeline. Second, four seed queries may not fully capture real user intent, which is often a latent construct not directly evident from surface phrasing; findings should be interpreted as evidence about these specific framings rather than user intent more broadly. Third, simulated behavior may not fully reflect realistic scrolling---the 7-second baseline is an unavoidable artifact of controlled auditing. Finally, fresh accounts and a 7-day window limit generalizability to mature profiles and longer-term dynamics.

\section{Discussion and Conclusion}

This study shows that in TikTok's mental health recommendations, post-search interaction behavior outweighs initial search framing in shaping what users ultimately see. Distress expression and explicit help-seeking are treated as broadly equivalent entry points into a shared mental health interest space, with differences in search intent reflected mainly in content mix rather than overall exposure. This suggests that both framings act as signals in the same semantic space, with similar recommendations reflecting embedding proximity.

A key implication is that help-seeking does not operate as a protective signal. While help-initiated searches shift recommendations toward a higher proportion of supportive material, potentially harmful content remains present---particularly when users continue watching mental health videos. 
We do not suggest that distress expressions deserve less protection than help-seeking queries; neither framing receives intent-sensitive treatment.
At the same time, strategies that suppress volume, such as avoidance or passive observation, do not fully remove risk or erase the effects of an initial search, highlighting an asymmetry between amplification and attenuation in the recommendation process. 

More broadly, these findings point to a misalignment between how users communicate intent through search and how recommender systems operationalize it through engagement optimization. In sensitive domains such as mental health, this misalignment may expose users seeking support to mixed-valence content that includes both supportive and harmful material. Addressing this gap will likely require recommendation designs that treat help-seeking context as safety-relevant during ranking and exploration, rather than relying primarily on behavioral reinforcement signals.

This work is a controlled, multi-day TikTok audit that separates intent framing from downstream behavioral interaction signals, enabling direct comparisons of persistence, amplification, and content composition under matched conditions. The results open several research directions, including audits with greater external validity and evaluating intervention designs that treat help-seeking context as safety-relevant during ranking and exploration rather than only topical.

\section{Ethical Considerations}
We audit TikTok's mental health recommendations using simulated accounts and report aggregate exposure patterns; we do not redistribute videos, identifiers, trace-level logs, or code/data artifacts. To reduce misuse, we avoid releasing automation scripts that could reproduce harmful pathways and keep conclusions scoped to the audited conditions and time.

\section{Acknowledgments}
We gratefully acknowledge financial support from the Huo Family Foundation. We thank Anne S. Warlaumont for her thoughtful and constructive suggestions. 
Accepted at the Proceedings of the International AAAI Conference on Web and Social Media (ICWSM 2026).

\bibliography{Ref}

@inproceedings{Boeker2022Empirical,
    author = {Boeker, Maximilian and Urman, Aleksandra},
    title = {An Empirical Investigation of Personalization Factors on TikTok},
    year = {2022},
    publisher = {Association for Computing Machinery},
    address = {New York, NY, USA},
    url = {https://doi.org/10.1145/3485447.3512102},
    doi = {10.1145/3485447.3512102},
    booktitle = {Proceedings of the ACM Web Conference 2022},
    pages = {2298--2309},
    numpages = {12},
    location = {Virtual Event, Lyon, France},
    series = {WWW '22}
}

@inproceedings{Mosnar2025Revisiting,
    author = {Mosnar, Matej and Skurla, Adam and Pecher, Branislav and Tibensky, Matus and Jakubcik, Jan and Bindas, Adrian and Sakalik, Peter and Srba, Ivan},
    title = {Revisiting Algorithmic Audits of TikTok: Poor Reproducibility and Short-term Validity of Findings},
    year = {2025},
    publisher = {Association for Computing Machinery},
    address = {New York, NY, USA},
    url = {https://doi.org/10.1145/3726302.3730293},
    doi = {10.1145/3726302.3730293},
    booktitle = {Proceedings of the 48th International ACM SIGIR Conference on Research and Development in Information Retrieval},
    pages = {3357--3366},
    numpages = {10},
    location = {Padua, Italy},
    series = {SIGIR '25}
}

@inproceedings{Xue2025Towards,
    title = {Towards an Automated Framework to Audit Youth Safety on {TikTok}},
    author = {Xue, Linda and Corso, Francesco and Fontana, Nicolo and Liu, Geng and Ceri, Stefano and Pierri, Francesco},
    booktitle = {Proceedings of the Fourth Workshop on Bridging Human-Computer Interaction and Natural Language Processing (HCI+NLP)},
    month = nov,
    year = {2025},
    address = {Suzhou, China},
    publisher = {Association for Computational Linguistics},
    url = {https://aclanthology.org/2025.hcinlp-1.9/},
    doi = {10.18653/v1/2025.hcinlp-1.9},
    pages = {113--119},
}

@article{Mousavi2024Auditing, 
    title={Auditing Algorithmic Explanations of Social Media Feeds: A Case Study of TikTok Video Explanations}, 
    volume={18}, 
    url={https://ojs.aaai.org/index.php/ICWSM/article/view/31376}, 
    DOI={10.1609/icwsm.v18i1.31376}, 
    number={1}, 
    journal={Proceedings of the International AAAI Conference on Web and Social Media}, 
    author={Mousavi, Sepehr and Gummadi, Krishna P. and Zannettou, Savvas}, 
    year={2024}, 
    month={May}, 
    pages={1110--1122} 
}

@misc{Baumann2025Dynamics,
    title={Dynamics of Algorithmic Content Amplification on TikTok}, 
    author={Fabian Baumann and Nipun Arora and Iyad Rahwan and Agnieszka Czaplicka},
    year={2025},
    eprint={2503.20231},
    archivePrefix={arXiv},
    primaryClass={physics.soc-ph},
    url={https://arxiv.org/abs/2503.20231}, 
}

@misc{Vombatkere2024TikTok,
      title={TikTok and the Art of Personalization: Investigating Exploration and Exploitation on Social Media Feeds}, 
      author={Karan Vombatkere and Sepehr Mousavi and Savvas Zannettou and Franziska Roesner and Krishna P. Gummadi},
      year={2024},
      eprint={2403.12410},
      archivePrefix={arXiv},
      primaryClass={cs.SI},
      url={https://arxiv.org/abs/2403.12410}, 
}

@misc{Eltaher2025Protecting,
    title={Protecting Young Users on Social Media: Evaluating the Effectiveness of Content Moderation and Legal Safeguards on Video Sharing Platforms}, 
    author={Fatmaelzahraa Eltaher and Rahul Krishna Gajula and Luis Miralles-Pechuán and Patrick Crotty and Juan Martínez-Otero and Christina Thorpe and Susan McKeever},
    year={2025},
    eprint={2505.11160},
    archivePrefix={arXiv},
    primaryClass={cs.SI},
    url={https://arxiv.org/abs/2505.11160}, 
}

@inproceedings{de2016discovering,
  title={Discovering shifts to suicidal ideation from mental health content in social media},
  author={De Choudhury, Munmun and Kiciman, Emre and Dredze, Mark and Coppersmith, Glen and Kumar, Mrinal},
  booktitle={Proceedings of the 2016 CHI conference on human factors in computing systems},
  pages={2098--2110},
  year={2016}
}

@techreport{narayanan2023understanding,
  author      = {Narayanan, Arvind},
  title       = {Understanding Social Media Recommendation Algorithms},
  institution = {Knight First Amendment Institute at Columbia University},
  series      = {Algorithmic Amplification and Society},
  year        = {2023},
  month       = apr,
  date        = {2023-04-26},
  doi         = {10.7916/khdk-m460},
  url         = {https://doi.org/10.7916/khdk-m460},
  note        = {Essay}
}

@article{krauth2025breaking,
  title={Breaking feedback loops in recommender systems with causal inference},
  author={Krauth, Karl and Wang, Yixin and Jordan, Michael},
  journal={ACM Transactions on Recommender Systems},
  volume={4},
  number={1},
  pages={1--20},
  year={2025},
  publisher={ACM New York, NY}
}

@inproceedings{coppersmith2016exploratory,
  title={Exploratory analysis of social media prior to a suicide attempt},
  author={Coppersmith, Glen and Ngo, Kim and Leary, Ryan and Wood, Anthony},
  booktitle={Proceedings of the third workshop on computational linguistics and clinical psychology},
  pages={106--117},
  year={2016}
}

@article{Naslund2016future, 
    title={The future of mental health care: peer-to-peer support and social media}, 
    volume={25}, 
    DOI={10.1017/S2045796015001067}, 
    number={2}, 
    journal={Epidemiology and Psychiatric Sciences}, 
    author={Naslund, J. A. and Aschbrenner, K. A. and Marsch, L. A. and Bartels, S. J.}, 
    year={2016}, 
    pages={113--122}
}

@article{Akhther2022SeekingSharing,
  author    = {Akhther, Najma and Sopory, Pradeep},
  title     = {Seeking and Sharing Mental Health Information on Social Media During COVID-19: Role of Depression and Anxiety, Peer Support, and Health Benefits},
  journal   = {Journal of Technology in Behavioral Science},
  year      = {2022},
  volume    = {7},
  number    = {2},
  pages     = {211--226},
  doi       = {10.1007/s41347-021-00239-x},
  url       = {https://doi.org/10.1007/s41347-021-00239-x}
}

@inproceedings{Milton2023ISee,
author = {Milton, Ashlee and Ajmani, Leah and DeVito, Michael Ann and Chancellor, Stevie},
title = {``I See Me Here'': Mental Health Content, Community, and Algorithmic Curation on TikTok},
year = {2023},
publisher = {Association for Computing Machinery},
address = {New York, NY, USA},
url = {https://doi.org/10.1145/3544548.3581489},
doi = {10.1145/3544548.3581489},
booktitle = {Proceedings of the 2023 CHI Conference on Human Factors in Computing Systems},
articleno = {480},
numpages = {17},
location = {Hamburg, Germany},
series = {CHI '23}
}

@article{Saha2020Causal, 
    title={Causal Factors of Effective Psychosocial Outcomes in Online Mental Health Communities}, 
    volume={14}, 
    url={https://ojs.aaai.org/index.php/ICWSM/article/view/7326}, 
    DOI={10.1609/icwsm.v14i1.7326}, 
    number={1}, 
    journal={Proceedings of the International AAAI Conference on Web and Social Media}, 
    author={Saha, Koustuv and Sharma, Amit}, 
    year={2020}, 
    month={May}, 
    pages={590--601} 
}

@Article{Gillian2025Self,
author="Grant-Allen, Gillian
and Wang, Lezhi
and Amini, Jasmine
and Dhaliwal, Simran
and Sinyor, Mark
and Mitchell, Rachel HB",
title="Self-Harm and Suicide-Related Content on TikTok: Thematic Analysis",
journal="J Med Internet Res",
year="2025",
month="Sep",
day="18",
volume="27",
pages="e77828",
issn="1438-8871",
doi="10.2196/77828",
url="https://doi.org/10.2196/77828",
}

@inproceedings{mansoury2020feedback,
author = {Mansoury, Masoud and Abdollahpouri, Himan and Pechenizkiy, Mykola and Mobasher, Bamshad and Burke, Robin},
title = {Feedback Loop and Bias Amplification in Recommender Systems},
year = {2020},
publisher = {Association for Computing Machinery},
address = {New York, NY, USA},
url = {https://doi.org/10.1145/3340531.3412152},
doi = {10.1145/3340531.3412152},
booktitle = {Proceedings of the 29th ACM International Conference on Information \& Knowledge Management},
pages = {2145–-2148},
numpages = {4},
location = {Virtual Event, Ireland},
series = {CIKM '20}
}

@article{Ribeiro2023Amplification, 
title={The Amplification Paradox in Recommender Systems}, 
volume={17}, 
url={https://ojs.aaai.org/index.php/ICWSM/article/view/22223}, 
DOI={10.1609/icwsm.v17i1.22223}, 
number={1}, 
journal={Proceedings of the International AAAI Conference on Web and Social Media}, 
author={Horta Ribeiro, Manoel and Veselovsky, Veniamin and West, Robert}, 
year={2023}, 
month={Jun.}, 
pages={1138--1142} 
}

@article{Franklin2017RiskFactors,
  author  = {Franklin, Joseph C. and Ribeiro, Jessica D. and Fox, Kathryn R. and Bentley, Khadijah H. and Kleiman, Evan M. and Huang, Xieyining and Musacchio, Kathryn M. and Jaroszewski, Amy C. and Chang, Benjamin P. and Nock, Matthew K.},
  title   = {Risk Factors for Suicidal Thoughts and Behaviors: A Meta-Analysis of 50 Years of Research},
  journal = {Psychological Bulletin},
  year    = {2017},
  volume  = {143},
  number  = {2},
  pages   = {187--232},
  doi     = {10.1037/bul0000084}
}

@article{Nesi2021SocialMediaSITB,
  author  = {Nesi, Jacqueline and Burke, Taylor A. and Bettis, Alexandra H. and Kudinova, Anastacia Y. and Thompson, Elizabeth C. and MacPherson, Heather A. and Fox, Kara A. and Lawrence, Hannah R. and Thomas, Sarah A. and Wolff, Jennifer C. and Altemus, Melanie K. and Soriano, Sheiry and Liu, Richard T.},
  title   = {Social Media Use and Self-Injurious Thoughts and Behaviors: A Systematic Review and Meta-Analysis},
  journal = {Clinical Psychology Review},
  year    = {2021},
  volume  = {87},
  pages   = {102038},
  doi     = {10.1016/j.cpr.2021.102038},
}

@article{Choudhury2014Mental, 
    title={Mental Health Discourse on reddit: Self-Disclosure, Social Support, and Anonymity}, 
    volume={8}, 
    url={https://ojs.aaai.org/index.php/ICWSM/article/view/14526}, 
    DOI={10.1609/icwsm.v8i1.14526}, 
    number={1}, 
    journal={Proceedings of the International AAAI Conference on Web and Social Media}, 
    author={De Choudhury, Munmun and De, Sushovan}, 
    year={2014}, 
    month={May}, 
    pages={71--80} 
}

@article{Nguyen2025Supporters, 
    title={Supporters and Skeptics: LLM-Based Analysis of Engagement with Mental Health (Mis)Information Content on Video-Sharing Platforms}, 
    volume={19}, 
    url={https://ojs.aaai.org/index.php/ICWSM/article/view/35875}, 
    DOI={10.1609/icwsm.v19i1.35875},  
    number={1}, 
    journal={Proceedings of the International AAAI Conference on Web and Social Media}, 
    author={Nguyen, Viet Cuong and Jain, Mini and Chauhan, Abhijat and Soled, Heather Jamie and Lesmes, Santiago Alvarez and Li, Zihang and Birnbaum, Michael L. and Tang, Sunny X. and Kumar, Srijan and De Choudhury, Munmun}, 
    year={2025}, 
    month={Jun.}, 
    pages={1329--1345} 
}

\appendix

\counterwithout{equation}{section}
\renewcommand{\theequation}{S\arabic{equation}}
\setcounter{equation}{0}
\renewcommand{\thetable}{S\arabic{table}}
\setcounter{table}{0}
\renewcommand{\thefigure}{S\arabic{figure}}
\setcounter{figure}{0}

\section{Appendix Section A: LLM Classification Prompt Structure}
\label{appendix:prompt}

The first-stage real-time classifier and the second-stage post-hoc labeler both use GPT-4o-mini with a structured prompt. The first stage receives a screenshot and extracts textual metadata, then returns a binary relevance decision to guide agent behavior; the second stage operates offline on the same extracted metadata to assign stance and harm-subtype labels. Table~\ref{tab:prompt-schema} summarizes the output schema, Table~\ref{tab:mh-criteria} lists the mental-health relevance criteria, and Table~\ref{tab:harm-categories} defines the harmful content subcategories used in downstream analysis.

\begin{table}[h]
\footnotesize
\renewcommand{\arraystretch}{1}
\begin{tabularx}{\columnwidth}{@{} l X @{}}
\toprule
\textbf{Field} & \textbf{Description} \\
\midrule
\texttt{creator\_username}          & Visible TikTok handle \\
\texttt{display\_name}              & Creator display name \\
\texttt{caption}                    & Main video caption \\
\texttt{hashtags}                   & Hashtags without \# symbol \\
\texttt{on\_screen\_text}           & Text overlaid on video \\
\texttt{language}                   & ISO language code (e.g., \texttt{en}) \\
\texttt{visual\_description}        & 2--3 sentence description: subject, setting, activity, tone \\
\texttt{possible\_mh\_relevance}    & True if mental-health relevant (see Table~\ref{tab:mh-criteria}) \\
\texttt{mental\_health\_keywords}   & Specific MH terms found \\
\texttt{mental\_health\_confidence} & \texttt{high} / \texttt{medium} / \texttt{low} \\
\texttt{content\_categories}        & Safety flags (see Table~\ref{tab:harm-categories}) \\
\bottomrule
\end{tabularx}
\caption{LLM prompt output schema. Each field is extracted from
         captions, hashtags, on-screen text, and visual elements
         visible in a single screenshot.}
\label{tab:prompt-schema}
\end{table}

\begin{table}[ht]
\centering
\footnotesize
\renewcommand{\arraystretch}{1}
\begin{tabularx}{\columnwidth}{@{} l X @{}}
\toprule
\textbf{Signal type} & \textbf{Examples} \\
\midrule
Explicit diagnoses     & Depression, anxiety, PTSD, bipolar disorder,
                         eating disorders, OCD, schizophrenia \\
\addlinespace
Crisis / self-harm     & Suicidal ideation, self-harm, suicide prevention,
                         crisis helplines (e.g., 988) \\
\addlinespace
Treatment terms        & Therapy, counseling, psychiatry, antidepressants,
                         mental health medication \\
\addlinespace
Emotional distress     & ``I want to die'', ``feeling hopeless'',
                         ``panic attack'', ``worthless'', ``alone'' \\
\addlinespace
Recovery language      & ``healing'', ``it gets better'', ``day \textit{N}
                         of recovery'', peer support \\
\addlinespace
Hashtags               & \#depression, \#mentalhealth, \#anxiety,
                         \#therapy, \#recovery, \#sad \\
\addlinespace
Visual cues            & Crying / distressed subject, therapy-office setting,
                         crisis resources displayed on screen \\
\addlinespace
\textit{Excluded}      & \textit{General wellness without MH framing
                         (e.g., yoga for relaxation, nutrition only)} \\
\bottomrule
\end{tabularx}
\caption{Mental-health relevance criteria used to set
         \texttt{possible\_mh\_relevance}. The classifier is instructed
         to prefer \textit{inclusion} under uncertainty.}
\label{tab:mh-criteria}
\end{table}

\begin{table}[t]
\centering
\footnotesize
\renewcommand{\arraystretch}{1}
\begin{tabularx}{\columnwidth}{@{} l X @{}}
\toprule
\textbf{Subcategory} & \textbf{Operational definition} \\
\midrule
Suicide / Self-Harm  & Explicit depiction of or instruction related to
                       suicidal behavior or self-injury; graphic imagery;
                       glorification of self-harm acts \\
\addlinespace
Toxic Positivity     & Dismissive ``good vibes only'' framing that
                       delegitimizes clinical distress or discourages
                       professional help-seeking \\
\addlinespace
Self-Diagnosis       & User-generated content diagnosing mental illness
                       without clinical basis; misleading symptom checklists
                       presented as definitive \\
\addlinespace
Misinformation       & Factually inaccurate claims about mental health
                       causes, treatments, or medication; content
                       contradicting evidence-based guidelines \\
\bottomrule
\end{tabularx}
\caption{Potentially harmful content subcategories used in
         second-stage post-hoc labeling.}
\label{tab:harm-categories}
\end{table}


\section{Appendix Section B: Extended Validation Results}
\label{appendix:ablation}

The F1 scores reported in the main text (MH-relevance $= 77.42\%$; potentially harmful $= 64.52\%$; potentially supportive $= 64.00\%$) are based on a random sample of 100 videos (${\simeq}1\%$ of the corpus), independently annotated by a human rater using the same taxonomy. Per-subcategory metrics for the four harmful content types are not reported separately due to the small validation sample size. 

\paragraph{Ablation study}

Multimodal inputs are necessary for real-time classification.
We test whether the real-time classifier's relevance decision is robust to removing perception channels (Table~\ref{tab:ablation_mh}). The full multimodal configuration attains the highest overall F1 and is the only setting that combines high recall with reasonable precision, prioritizing coverage in a safety-critical setting. Removing visual descriptions or on-screen text produces modest drops in F1 but large drops in recall, with corresponding gains in precision, indicating a systematic trade-off between coverage and conservativeness. When all visual channels are removed and the classifier relies only on captions and hashtags, F1 is roughly halved and precision collapses despite recall remaining relatively high. These patterns indicate that no single modality is sufficient for reliable real-time relevance detection: visual information and extracted on-screen text are crucial for avoiding missed mental-health content, while textual cues help refine precision. This supports our design choice to use multimodal perception for first-stage real-time classification, while reserving richer categorization for second-stage post-hoc labeling.

\begin{table*}[t]
\centering
\footnotesize
\renewcommand{\arraystretch}{1.29}
\begin{tabular}{p{6.2cm} p{7.05cm} p{0.7cm} p{0.9cm} p{0.8cm}}
\hline
\textbf{Approach} & \textbf{What Model Sees} & \textbf{F1} & \textbf{Precision} & \textbf{Recall} \\
\hline
\textbf{Our Model (Full Multimodal)} &
Screenshots + visual\_description + on\_screen\_text + caption + hashtags &
77.42\% & 63.16\% & 100\% \\

Vision, No Visual Desc &
Screenshots + on\_screen\_text + caption + hashtags &
73.7\% & 84.8\% & 65.1\% \\

Vision, No On-Screen &
Screenshots + visual\_description + caption + hashtags &
70.1\% & 79.4\% & 62.8\% \\

No Vision [Text] &
visual\_description + on\_screen\_text + caption + hashtags &
61.8\% & 48.8\% & 84\% \\

No Vision, No On-Screen [Text] &
visual\_description + caption + hashtags &
57.6\% & 44.2\% & 82.6\% \\

No Vision, No Visual Desc [Text] & on\_screen\_text + caption + hashtags &54\% & 39.5\% & 85\% \\

No Vision, No On-Screen, No Visual Desc [Text] & caption + hashtags & 35.7\% & 23.3\% & 76.9\% \\
\hline
\end{tabular}
\caption{Ablation study results for binary mental-health relevance detection.}
\label{tab:ablation_mh}
\end{table*}

\section{Appendix Section C: Operational Reliability}
\label{appendix:ops}
We assess whether the pipeline operates reliably across multi-day, multi-account collection (Table~\ref{tab:ops}). Classifier latency fits within the baseline viewing window (median 5.2\,s), and the second-stage post-hoc labeler is substantially faster, with typical annotations completed in under a second. All planned sessions were completed successfully, and all collected videos were processed by the analysis pipeline. No session-level classifier failures occurred; occasional errors triggered a conservative fallback that defaulted to skip after $t_{\text{base}}$ ($\simeq$7 seconds) while still logging a complete trace. During collection, one account was suspended mid-study and two session cookies expired; cookies were renewed without disrupting the routine, and the suspension reduced the number of account-days but did not affect trace integrity for completed sessions. Together, these results indicate the pipeline supports sustained audits with reproducible traces and bounded operational overhead.

\begin{table}[h]
\centering
\footnotesize
\begin{tabular}{lr}
\toprule
\textbf{Metric} & \textbf{Value} \\
\midrule
\multicolumn{2}{l}{\textit{Controller (LLM) latency}} \\
\quad Median (p50) & 5.20\,s \\
\quad 95th percentile (p95) & 6.04\,s \\
\quad 99th percentile (p99) & 6.42\,s \\
\midrule
\multicolumn{2}{l}{\textit{Labeler (Vision API) latency} ($n=120$ videos, sampled)} \\
\quad Median (p50) & 0.48\,s \\
\quad 95th percentile (p95) & 0.77\,s \\
\quad 99th percentile (p99) & 0.92\,s \\
\midrule
\multicolumn{2}{l}{\textit{Controller reliability} ($n=138$ sessions)} \\
\quad Session abort rate & 0.0\% (0/138) \\
\quad Sessions with $\geq$1 fallback & 29.0\% (40/138) \\
\midrule
\multicolumn{2}{l}{\textit{End-to-end completion} 
(all 30 accounts)} \\
\quad Session completion rate & 100.0\% (208/208) \\
\quad Video analysis rate & 100.0\% (8727/8727) \\
\quad Avg.\ videos per session & 42.0 \\
\bottomrule
\end{tabular}
\caption{Operational reliability metrics. Latencies are wall-clock. ``Fallback'' denotes default action ($t_{\text{base}}$) used when classifier output was unavailable or invalid; ``failure'' denotes session abort (no completed trace).}
\label{tab:ops}
\par\smallskip
\begin{minipage}{\columnwidth}
\footnotesize
\textit{Note:} $n=138$ covers MH-Engaged and MH-Avoidant only; Passive-Observer agents extract metadata but apply fixed watch durations. $n=208$ covers all 30 accounts.
\end{minipage}
\end{table}

\section{Appendix Section D: Exposure Trajectories and Descriptive Statistics}
\label{appendix:suppfig}

\begin{figure*}[t]
    \centering
    \includegraphics[width=1\linewidth]{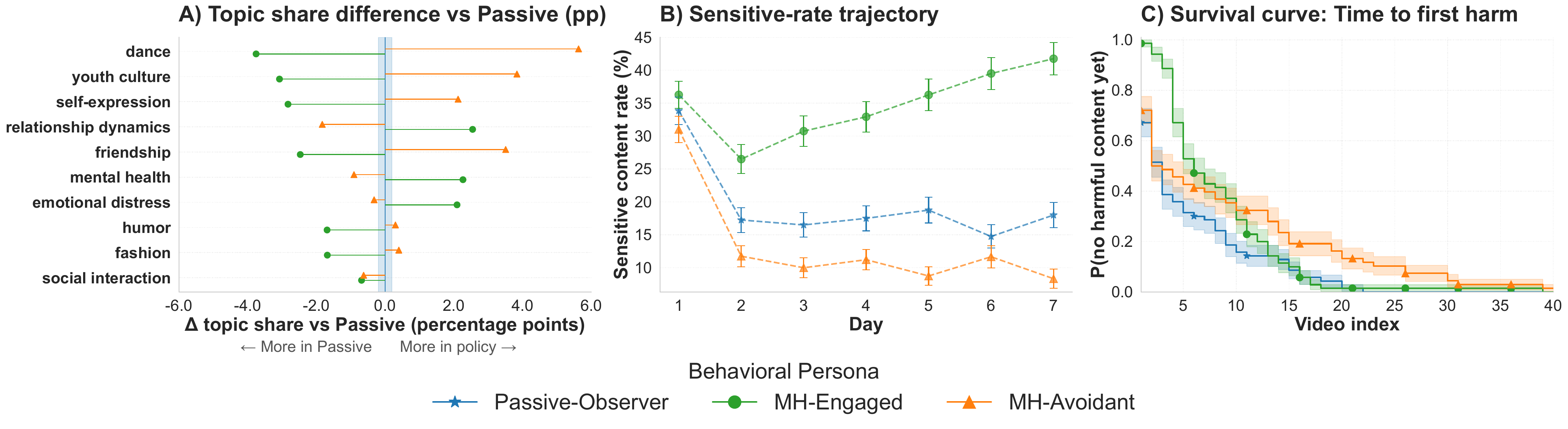}
    \caption{\textbf{Exposure trajectories across behavioral personas.}
    (A) Topic share differences vs Passive (percentage points).
    (B) Sensitive-content rate over time.
    (C) Survival curve: probability of no harmful content as a function of video index. (Lower curve = earlier harmful exposure)}    
    \label{fig:suppfig}
\end{figure*}

Figure~\ref{fig:suppfig} shows that behavioral interaction policy induces directional shifts in both feed composition and exposure timing, confirming that interaction policy alone---independent of initial search framing---is sufficient to produce distinct exposure trajectories under matched session budgets.

\textbf{Topic composition (Figure~\ref{fig:suppfig}A).} Relative to Passive-Observer, MH-Engaged shifts the feed toward mental health ($+$2.27\,pp) and emotional distress ($+$2.09\,pp), with corresponding reductions in lighter topics such as dance ($-$3.76\,pp) and youth culture ($-$3.07\,pp). MH-Avoidant shows the mirror pattern: dance ($+$5.63\,pp) and youth culture ($+$3.84\,pp) increase while mental health is suppressed ($-$0.91\,pp), consistent with each policy's intent to reinforce or withhold reinforcement of MH-relevant content.

\textbf{Sensitive-content trajectory (Figure~\ref{fig:suppfig}B).} On Day 1, all personas begin at comparable rates ($\simeq$31--36\%), reflecting shared initialization from the seed search phase. From Day 2, policies diverge sharply: MH-Engaged rises steadily to $\simeq$42\% by Day 7, Passive-Observer stabilizes at $\simeq$15--19\%, and MH-Avoidant declines to $\simeq$8--12\% and remains stable. Sensitive content does not disappear under avoidance, indicating that a single MH-related search leaves a lasting footprint even in the absence of continued engagement---consistent with the main text findings.

\textbf{Time to first harmful exposure (Figure~\ref{fig:suppfig}C).} By video index 10, only 15.7\% of Passive-Observer sessions and 28.6\% of MH-Engaged sessions remain harm-free, compared to 32.4\% of MH-Avoidant sessions. By index 20, MH-Avoidant retains 13.2\% harm-free sessions versus near-zero for the other two. Across all personas, 98.5--100\% of sessions encountered at least one potentially harmful video by index 40, indicating that avoidance reduces but does not eliminate exposure to potentially harmful content---mirroring the ``avoidance reduces volume but not risk'' finding reported in the main text.

\textbf{Descriptive Statistics.}
Table~\ref{tab:descriptives} summarizes session-level exposure outcomes by policy and framing condition. Differences in potentially harmful exposure were large and consistent across framing: MH-Engaged accounts showed mean potentially harmful rates roughly twice those of MH-Avoidant accounts (22.5\% vs.\ 10.3\%), with Passive-Observer accounts intermediate (16.0\%). Help-Init.\ framing consistently yielded lower potentially harmful rates within each behavioral condition. FYP-specific potentially harmful rates were substantially elevated for MH-Engaged accounts (42.4\%) relative to Passive-Observer (15.0\%) and MH-Avoidant (8.5\%).

\begin{table*}[t]
\centering
\footnotesize
\small
\begin{tabular}{llccccc}
\toprule
Policy & Framing & $n$ & Mean Videos & Potentially Harmful Rate & MH Rate & FYP Potentially Harmful Rate \\
\midrule
\multirow{2}{*}{Passive-Observer}    & Distress-Init. & 35 & 41.8 & .189 (.084) & .202 (.102) & .163 (.106) \\
                             & Help-Init.     & 35 & 41.6 & .130 (.067) & .163 (.079) & .136 (.098) \\
\addlinespace
\multirow{2}{*}{MH-Engaged} & Distress-Init. & 35 & 42.2 & .272 (.089) & .453 (.162) & .419 (.213) \\
                             & Help-Init.     & 35 & 41.9 & .178 (.093) & .454 (.178) & .429 (.231) \\
\addlinespace
\multirow{2}{*}{MH-Avoidant}   & Distress-Init. & 35 & 42.2 & .117 (.065) & .126 (.064) & .093 (.046) \\
                             & Help-Init.     & 33 & 42.0 & .088 (.047) & .109 (.053) & .077 (.043) \\
\midrule
Overall & --- & 208 & 41.9 & .163 (.096) & .252 & .224 \\
\bottomrule
\end{tabular}
\caption{Descriptive Statistics by Policy and Framing Condition}
\label{tab:descriptives}
\par\smallskip
\raggedright\footnotesize
\textit{Note.} Values are means with SDs in parentheses. Potentially Harmful Rate = proportion of all session videos (including search phase) labeled potentially harmful. FYP Potentially Harmful Rate = proportion of FYP-only videos labeled potentially harmful; this is the measure reported in the main text and figures. MH Rate = proportion of all session videos labeled MH-relevant.
\end{table*}

\end{document}